\documentstyle[aps,epsf,preprint]{revtex}
\begin{document}
\draft
\preprint{Submitted to Physical Review B}

\title{Quasi One-Dimensional Spin Fluctuations in YBa$_2$Cu$_3$O$_{6+x}$}
\author{D. N. Aristov$^{1,2}$ and D. R. Grempel$^1$}
\address{${}^1$CEA/D\'epartement de Recherche Fondamentale sur
la Mati\`ere Condens\'ee}
\address{SPSMS/MDN, CENG, 17, rue des Martyrs, 38054 Grenoble
Cedex 9, France}
\address{${}^2$Petersburg Nuclear Physics Institute, 188350
Gatchina, Saint-Petersburg, Russia}

\date{\today}

\maketitle

\begin{abstract}
We study the spin fluctuation of the oxygen deficient
planes of YBa$_2$Cu$_3$O$_{6+x}$. The Cu-O chains that
constitute these planes are described by a model that includes
antiferromagnetic interactions between  spins and Kondo-like
scattering of oxygen holes. The spectrum of magnetic excitations
shows the presence of incommensurate dynamic fluctuations along
the direction of the chains. The presence of itinerant holes is responsible for the existence of important differences between the spin dynamics
of this system and that of a quasi-one-dimensional localized antiferromagnet. We comment on the
possibility of experimental observation of these fluctuations.
\end{abstract}


\narrowtext

\section{INTRODUCTION}
\label{sec:intro}
A considerable amount of experimental and theoretical research
has recently been consecrated to the study of the structural and
electronic properties of the oxygen deficient (or CuO$_x$)
planes of superconductors of the YBa$_2$Cu$_3$O$_{6+x}$-type. It
has been known for some time that these planes play an important
 role in the superconductivity of the YBCO compounds by
regulating the hole concentration in the CuO$_2$
planes\cite{Ui}. There is now evidence that they also have an
interesting physics of their own \cite{mook,bansov}. The basic
components of the CuO$_x$ planes are Cu$_{n+1}$O$_n$ chain
fragments in which Cu and O atoms form a linear alternating
chain. Under stoichiometric conditions, $x=0.5$ and $x=1$,  all
chains are ( ideally) infinitely long.  Away from stoichiometry
the length of the chains follows a certain distribution
determined by $x$ and the temperature but the physical
properties of the chain planes are dominated by the longest
chain fragments\cite{hohl,schleg}. Band structure
calculations\cite{Mass,Pick} for YBa$_2$Cu$_3$O$_7$ show that
the dispersion of the chain-derived bands which cross the Fermi
level in the directions perpendicular to the chains is much
weaker than that in the direction of the chains. This suggests
that, to a first approximation,  the latter may be viewed as
decoupled from each other and from the CuO$_2$ planes. If this
is correct, important $1\!-\!d$ fluctuation effects are expected
in the  charge and spin excitation spectra of the CuO$_x$
planes. A recent inelastic neutron scattering study\cite{mook}
of a YBa$_2$Cu$_3$O$_{6.93}$ crystal shows that such effects are
indeed visible in the phonon spectrum of the chain planes. Their
signature is the presence of purely inelastic satellites around
the structural Bragg peaks at positions that correspond closely
to the value of $2k_F$ for the one-dimensional chains 
 predicted by the band structure calculations\cite{Pick}.

In this work we study the spin fluctuation spectrum of long
Cu$_{n+1}$O$_n$ chains in the framework of a one-dimensional
model of strongly interacting fermions.  Most of the paper deals
with the $T\!=\!0$ case but we briefly discuss the extension of
our calculations to finite temperatures in the concluding
Section. Our main results are the following. The  spin-spin
correlation function of the chains is given by the convolution
of two functions. One is the dynamic structure factor of a
one-dimensional Heisenberg antiferromagnet on a fictitious
lattice that contains a number of sites equal to the total
number of holes in the chain. This factor describes the part
of the dynamic that results from the exchange interactions
between the spins in the chain. The second factor, related to
the density-density correlation function of the itinerant oxygen
holes, takes into account the perturbation of the spin configurations
that arise from the motion of the latter.  The magnetic
intensity is modulated along the direction of the chains
\cite{foot0} and is peaked at an incommensurate position,
$Q_{max}\!=\!\pi(1+c)$,  where $c$ is the  nominal oxygen hole
concentration in the chains (see below). The lineshape of the
energy-integrated peak, very asymmetric, is quite different from
that obtained for a  one-dimensional Heisenberg antiferromagnet.
The intensity at $Q_{max}$ is distributed over an energy range given
 by the Fermi energy of the holes. However, enough intensity
 remains at low energies that observation of these excitations
 by inelastic neutron scattering might be feasible.

The organization of the rest of the paper is as follows. The
model used to describe the Cu-O chains is presented in Section
\ref{sec:model}. The calculation of the  ground state spin-spin
correlation function is described in \ref{sec:corr}. In Section
\ref{sec:discussion} we discuss our results and their connection
with experiment.

\section{ THE MODEL.}
\label{sec:model}
The simplest microscopic model for the Cu-O chains is the
$1\!-\!d$ Emery Hamiltonian\cite{Em} defined by:

\begin{eqnarray}
\label{1}
H=-t\sum_{<r,\rho>}\sum_{\sigma}\left(p^{\dagger}_{r\sigma}d^{}_{\rho\sigma}+
d^{\dagger}_{\rho\sigma}p^{}_{r\sigma}\right)+
\sum_r(\epsilon_pn_r+ U_pn_{r\uparrow}n_{r\downarrow})+
\sum_{\rho}\left(\epsilon_dn_{\rho}+U_dn_{\rho\uparrow}n_{\rho\downarrow}
\right),
\end{eqnarray}
\vspace{-4mm}

\noindent
Here, $r$ and $\rho$ denote the O and Cu sites, respectively,
$d_{\sigma}$ and $p_{\sigma}$ are annihilation operators for
holes of spin $\sigma$ on Cu and O sites, the occupation number
of a hole with spin $\sigma$ is $n_{\sigma}$ and
$n\!=\!\sum_{\sigma}n_{\sigma}$. Hopping is only allowed between
nearest neighbors.  The values of the charge transfer gap and of
the on-site Coulomb repulsion  appropriate for the description
of the CuO chains are\cite{Gre-Ui} $(\epsilon_p-\epsilon_d
)=2t$, and $U_d=2 U_p=8 t$, where $t\approx 1 eV$ is the Cu-O
transfer integral. In the following we shall measure energies
from the position of the Cu level and set $\epsilon_d\!=\!0$.

An additional parameter, the charge of the fragment, has to be
fixed in order to completely define the problem. Taking the
Cu$^{+1}$ and O$^{-2}$ configurations as reference states, one
may easily  see\cite{Ui} that a neutral fragment with $N$ Cu
atoms contains $2N-1$ holes. However, because of the transfer of
charge between the CuO$_x$ and the CuO$_2$ planes the number of
holes in the chains, $N(1+c)$ with $0\le\!c\le\!1$,  deviates
from this value\cite{Gre-Ui}. In a naive picture in which all
the Cu ions in the chain are in the Cu$^{2+}$ state, $c$ is
simply the concentration of oxygen holes. It is experimentally
known \cite{krol} that the hole concentration in long chain
fragments is $\nu_h\!\approx\!1.3$. Therefore, for the systems
discussed here, $c\!  \approx\!0.3$. It is however interesting
to treat $c$ as a free parameter and study the  dependence on
concentration of different physical quantities.

The Emery Hamiltonian of Eq.\ \ref{1} gives a good a description
of the static electronic properties of the chain
fragments\cite{Gre-Ui,Gre-UiII} but it can only be properly
treated with numerical methods such as Quantum Monte Carlo or
exact diagonalization that are not well suited for the
calculation of the dynamic susceptibility. There is a limiting
case, $U_d/t\rightarrow \infty$, $\epsilon_p/t$ and $U_p/t$
finite but large, in which a more tractable model\cite{Go-Io}
can be derived from Eq.\ \ref{1}. In this limit the Cu
occupation $n_{ \rho}$ is unity and charge fluctuations on the
Cu sites only appear in virtual processes generated in
perturbation theory by the kinetic energy term. In the 2nd order
in $t$ we generate a process in which an O hole hops from a site
$r$ to an adjacent O site, $r{\pm}2$ leaving the spin {\it
sequence} in the lattice unchanged\cite{foot1}. This process,
whose amplitude is $\tau\!=\!t^2/\epsilon_p$, is described by a
tight-binding model for {\it spinless} free fermions defined on
the oxygen sublattice:

\begin{eqnarray}
\label{h1}
H_1=-\tau\sum_{j=1}^{N}(c^{\dag}_{j+1+1/2}c^{}_{j+1/2} + h.c.).
\end{eqnarray}

\noindent In the same order in $t$  there  is an exchange
interaction between a spin on an oxygen site and the spins on
neighboring Cu sites with coupling
 $J=t^2/\left(\epsilon_p+U_p\right)$.  Finally, if there is no O
hole on a site $r$ the Cu spins on sites $r\pm 1$ still interact
antiferromagnetically with a coupling constant that is of the
fourth-order in $t$, $J'=4 t^4/(\epsilon_p^2
(2\epsilon_p+U_p))$. The spin part of the interaction is thus
described by a $1\!-\!d$ Heisenberg Hamiltonian given by :

\begin{eqnarray}
\label{h2}
H_2=\sum_{i=1}^{N+N_h-1} J_{i,i+1} {\bbox \sigma}_i . {\bbox \sigma}_{i+1},
\end{eqnarray}

\noindent where $N_h\equiv N c$ is the number of oxygen holes in
the system,  $N+N_h$ is the total number of spins in the system
and ${\bbox \sigma}_i$ are spin 1/2 operators.  The
nearest-neighbor coupling $J_{i,i+1}\!=\!J$ or $J'$ depending
on whether $(i,i+1)$ is a Cu-O or a Cu-Cu pair, respectively.

Since the couplings $J_{i,i+1}$ depend upon the instantaneous
hole distribution, the spin and charge degrees of freedom of the
system are coupled and the total Hamiltonian is not exactly
solvable in general. However, in the special case $J'/J\!=\!1$
an exact solution can be found. This is because if the exchange
coupling is the same for all pairs of spins regardless of the
charge configuration,  the Hamiltonian becomes the sum of two
independent terms each of which is exactly solvable. In
the case of the Cu-O chains the ratio of couplings $J'/J\approx
0.75$ is not unity but, as shown elsewhere\cite{Gre-Ui}, their {\it static}
magnetic correlations are quite accurately described by an
effective homogeneous model.  It seems therefore natural to try and
use it to compute the time-dependent correlation functions as well. This is the approach that we follow in this paper. 

The appropriate parametrization of the effective model may be
determined variationally. If we construct a trial ground-state
wave function as the product of separate wavefunctions for the
spins and for the holes and then determine the individual
components by minimizing the total energy of the system, we
obtain the effective model

\begin{eqnarray}
\label{gi}
H_{eff}= -\tau\sum_{j=1}^{N}(c^{\dag}_{j+1+1/2}c^{}_{j+1/2} + {\rm
h.c.})+ J_{eff}(c) \sum_{i=1}^{N+N_h-1} {\bbox \sigma}_i . {\bbox
\sigma}_{i+1}.
\end{eqnarray}

\noindent where $J_{eff}(c)\!=\!J'(1-c)+c J$ and we have omitted
an energy shift that can be absorbed in a redefinition of the
chemical potential of the holes. We find $J_{eff}/\tau\approx
0.3$ for $c=0.3$.

In spite of the simple form of $H_{eff}$ the
spin-spin correlation function,

\begin{eqnarray}
\label{sqwdef}
{\cal S}(Q,\Omega)=\int_{-\infty}^{\infty} dt e^{i \Omega t}
\left\langle {\bf S}_Q(t).{\bf S}_{-Q}(0)\right\rangle,
\end{eqnarray}

\noindent is non-trivial because the relationship between
${\bf S}_i$ and $\sigma_i$ is not simple\cite{Vi-Bak}. The
difference between the coordinates of a spin measured in
the real and in the fictitious lattices is determined by the number
of holes located to its left. In terms of ${\bbox \sigma}_k$, the
Fourier transform of the spins on the fictitious lattice,  we
have:

\begin{eqnarray}
\label{spinop1}
{\bf S}_l=(N+N_h)^{-1/2} \sum_k^{} {\bbox \sigma}_k \exp\ i k
(l+\sum_{j=1}^{l-1} c_{j+1/2}^{\dagger} c_{j+1/2}^{}),
\end{eqnarray}

\noindent
for Cu sites, and

\begin{eqnarray}
\label{spinop2}
{\bf S}_{l+1/2}=(N+N_h)^{-1/2} \sum_k^{} {\bbox \sigma}_k \exp\ i
k ( l+1+\sum_{j=1}^{l-1} c_{j+1/2}^{\dagger} c_{j+1/2}^{}) \
c_{l+1/2}^{\dagger} c_{l+1/2}^{},
\end{eqnarray}

\noindent
for O sites that only carry a spin when they are occupied by a hole.

\noindent Substituting Eqs.\ \ref{spinop1} and \ref{spinop2} in
Eq.\ \ref{sqwdef} and using the fact that the ground-state
wavefunction of the decoupled Hamiltonian factorizes,  ${\cal
S}(Q,\Omega)$ may be written as a convolution  :

\begin{eqnarray}
\label{sqw}
{\cal S}(Q,\Omega)=\int_{-\pi}^{\pi} \frac{dk}{2\pi}
\int_{-\infty}^{\infty} \frac{d\omega}{2\pi}
|A(Q,k)|^2 f(k,\omega)\  {\cal
C}_k (Q-\tilde{k},\Omega-\omega),
\end{eqnarray}

\noindent where $\tilde{k}=k(1+c)$.

\noindent The first factor in Eq.\ \ref{sqw}, $f(k,\omega)$, is
the dynamic spin correlation function of the fictitious
$1\!-\!d$ Heisenberg model of Eq.\ \ref{h2},

\begin{eqnarray}
\label{heisen}
f(k,\omega)=\int_{-\infty}^{\infty} dt\,e^{i \omega t}\left\langle
{\bbox \sigma}_{k}(t).{\bbox \sigma}_{-k}(0)\right\rangle_{H_2}.
\end{eqnarray}

\noindent The second factor, ${\cal C}_k$, describes the effects
of the hole dynamics and is given by the space-time Fourier
transform of :

\begin{eqnarray}
\label{fermionic}
{\cal C}_k(l,l';t,t')=\left\langle \exp\left[ i k \sum_{q \ne 0}
\phi(q,l) \rho(q,t)\right] \exp\left[ -i k \sum_{q \ne 0}
\phi(q,l') \rho(q,t')\right] \right\rangle_{H_1},
\end{eqnarray}

\noindent
where the density operator is given by :

\begin{eqnarray}
\label{dsty}
\rho(q)= N^{-1} \sum_p c^{\dagger}_{p} c^{}_{p+q},
\end{eqnarray}

\noindent
and we have defined $\phi (l,q)=(1-e^{i q l})/(1-e^{i q})$. The
$q\!=\!0$ component of the density is not present in Eq.\
\ref{fermionic}. Searate treatment of this component leads to the appearance of $\tilde{k}$ instead of $k$ in the argument of ${\cal C}_k$ in Eq.\ \ref{sqw}. Finally, $A(Q,k)$ is a coherence factor given by
\begin{eqnarray}
\label{coherence}
A(Q,k)=1+\frac{\sin (Q-k)/2}{\sin(k/2)}.
\end{eqnarray}

\noindent This factor takes into account the phase
shift between Cu and O spins and it is derived in Appendix
\ref{sec:app:coh}.

\section{EVALUATION OF THE CORRELATION FUNCTION.}
\label{sec:corr}
The evaluation of the dynamic structure factor in Eq.\ \ref{sqw}
requires the  knowledge of the individual correlation functions
$f(k,\omega)$ and ${\cal C}_k(q,\omega)$. The dynamic
correlation function of the one-dimensional Heisenberg
antiferromagnet has been discussed in reference\cite{MTBB}. This
function cannot be obtained in closed form but a very accurate
approximate expression for it has been derived\cite{MTBB} using
a combination of exact analytic and numerical results :

\begin{eqnarray}
\label{heisapp}
f(k,\omega)=3 a \ \theta(\omega) \ \theta\left(\pi J_{eff}
\sin(|k|/2)-\omega\right) \
\left(\omega^2-(\frac\pi2 J_{eff}\sin k)^2\right)^{-1/2},
\end{eqnarray}

\noindent where $a=1.4579$ is a numerical constant. There are no
sharp spinwave excitations in the system but  the spectral
function is strongly peaked at the lower edge of the spectrum
where the spectral weight diverges as
$(\omega-\omega_s(k))^{-1/2}$ with $\omega_s(k)=\pi
J_{eff}/2\sin k$.

It will be seen below that most of the spectral weight in ${\cal
S}(Q,\Omega)$ is concentrated in the neighborhood of $Q_{max}\!=\!\pi
(1+c)$ (mod $2\pi$). Since $f(k,\omega)$ is strongly peaked, the structure factor in the neighborhood of $Q_{max}$ is determined by the long distance part
of the fermionic correlator ${\cal C}_k$ in Eq.  \ref{sqw}.  
This part of the response is
determined by the low energy excitations of the electron gas in
which the momenta of excited particles and holes are all near
$\pm k_F$,  the Fermi momentum of the holes.

Eq.\ \ref{fermionic} may be evaluated using a well known procedure\cite{Lu-Pe} in which the fermion density is split into
contributions from right and left moving fermions with constant
group-velocity. Since the fermionic Hamiltonian is particle-hole
symmetrical it is sufficient to compute ${\cal C}_k$ for
$c\!\leq\!1/2$ and make the transformation $c\rightarrow (1-c)$
to obtain the corresponding expression for $c>1/2$. We write

\begin{eqnarray}
\label{defrho}
\rho(q)=\rho_{+}(q)+\rho_{-}(q)\equiv N^{-1} \sum_{p>0}
c^{\dagger}_{p} c^{}_{p+q}+ N^{-1}\sum_{p<0} c^{\dagger}_{p}
c^{}_{p+q},
\end {eqnarray}

\noindent and approximate the group velocity of right and left
moving fermions by $\pm v_F$ respectively, where the Fermi
velocity is $v_F\!=\!2\tau \sin k_F$ and $k_F\!=\!\pi c$
($k_F\!=\!\pi (1-c)$ for $c>1/2$) . With this approximation the
time dependence of the density operators becomes very simple, $
\rho_{\pm}(q,t)= e^{\pm i\omega_qt}\rho_{\pm}(q)$ and
$\omega_q=v_F\left|q\right|$.

The long wave-length density fluctuations of the
$1\!-\!d$ electron gas may be represented by boson operators
defined by the relations\cite{Ma-Li} :

\begin{eqnarray}
\label{bosons}
\begin{array}{l}
\rho_{+}(q)=v(q) \left(\theta(q) b_{q}^{+}+\theta(-q) b_{-q}^{ }\right)
\\
\rho_{-}(q)=v(q) \left(\theta(q) b_{-q}^{ }+\theta(-q) b_{q}^{+}\right),
\end{array}
\end{eqnarray}

\noindent where $v(q)=\sqrt{|q|/(2N\pi)}$. With the definition

\begin{eqnarray}
\label{defa}
{\cal A}(l,t)\equiv \sum_{q} \phi(q,l) \rho(q,t)=\sum_{q} v(q)
\phi(q,l) \left[e^{i \omega_q t} b_{q}^{+}+e^{-i \omega_q t}
b_{-q}^{ }\right],
\end{eqnarray}

\noindent we may rewrite Eq.\ \ref{fermionic} in the form :

\begin{eqnarray}
\label{defb}
{\cal C}_k(l,l';t,t')=\left<\exp ik{\cal A}(l,t) \ \exp -ik
{\cal A}(l',t')\right>_{H_1}
\end{eqnarray}

The evaluation of this expression in Appendix
\ref{sec:app:fermionic} yields :

\begin{eqnarray}
\label{corrfermions}
{\cal C}_k(l,l';t,t')=\exp \left[-\frac{k^2}{4 \pi^2} {\cal
F}(l,l',t-t')\right],
\end{eqnarray}

\noindent where

\begin{eqnarray}
\label{decay}
{\cal F}(l,l';t)=4 \pi^2 \sum_{|q|\le k_F} v^{2}(q)
\frac{e^{-i\omega_q t} (1-\cos q(l-l'))+(1-e^{-i\omega_q t})
\left(2-\cos ql-\cos ql'\right)}{2(1-\cos q)}.
\end{eqnarray}

The integration over momenta in Eq. \ref{decay} is cutoff at $q\sim k_F$, the wave vector below which the density fluctuations of the electron gas can be treated as well defined excitations. In the thermodynamic limit ($l,l'\rightarrow
\infty, |l-l'|$ finite ) ${\cal F}(l,l';t)$ depends on $l$ and
$l'$ only through the difference $(l-l')$. Evaluating the
integral over momentum in Eq. \ref{decay} in the logarithmic
approximation we find :

\begin{eqnarray}
\label{defc}
{\cal C}_k(l;t)=\left[(1+ik_F(l+v_Ft)) (1-ik_F(l-v_Ft)) (1+iE_F
t)^{2}\right]^{-\alpha_k},
\end{eqnarray}

\noindent where $E_F=v_F k_F$ is the Fermi energy, and
$\alpha_k=(k/2\pi)^2 \le 1/4$.

The space-time Fourier transform of ${\cal C}_k(l;t)$ can be
found analytically. For $q$ in the first Brillouin zone :

\begin{eqnarray}
\label{cqw}
{\cal C}_k(q,\omega)= \frac{4\pi^2 \theta (\omega)\ (2
k_F)^{1-4\alpha_k}\  e^{-\omega/E_F}}{E_F \ \Gamma(\alpha_k)
\Gamma(3\alpha_k) \ (\omega^2/v_F^2-q^2 )^{1-3\alpha_k}
|q|^{2\alpha_k}}
F(\alpha_k,2
\alpha_k-1/2,3\alpha_k;1-\left(\omega/qv_F\right)^2),
\end{eqnarray}

\noindent where $F$ is the Gauss hypergeometric function. ${\cal
C}_k(q,\omega)$ diverges at the lower end of the spectrum, at
$\omega\!=\!|q|v_F$, the dispersion relation for density waves
in the electron gas. At high energies the spectrum is cut off
exponentially at $\omega \gtrsim E_F$.

The instantaneous correlation functions can be obtained by
integrating Eqs.\ \ref{heisapp} and \ref{cqw} over frequency.
For the Heisenberg model the result is\cite{MTBB} :

\begin{eqnarray}
\label{fstatic}
f(k)=\frac{3a}{2\pi}\
\log\left|\frac{1+\sin|k/2|}{\cos(k/2)}\right| \approx
\frac{3a}{2\pi}  \log\left|\frac{4}{\pi-|k|}\right|.
\end{eqnarray}

\noindent The approximate expression quoted on the right hand
side of Eq.\ \ref{fstatic} derived near the zone boundary at
$k\!=\!\pm\pi$ where $f(k)$ diverges logarithmically is in fact
remarkably accurate over most of the Brillouin zone and will be
used in our calculations below. The instantaneous value of the
fermionic correlator is  given by :

\begin{eqnarray}
\label{cstatic}
{\cal C}_k(q)=\frac{2^{1-\alpha_k}\
\sqrt{2\pi}}{\Gamma(\alpha_k)\ k_F}\left|\frac{q}{k_F}\
\right|^{-(1/2-\alpha_k)}K_{1/2-\alpha_k}(q/k_F).
\end{eqnarray}

\noindent ${\cal C}_k(q)$ is strongly peaked at $q\!=\!0$ where
it diverges as ${\cal C}_k(q)\approx q^{-(1-2\alpha_k)}$.  Away
from the center of the Brillouin zone ${\cal C}_k(q)$ decays
exponentially,  ${\cal C}_k(q)\approx \exp(-|q|/k_F)$.

The rapid decrease of $f(k,\omega)$ away
from the zone boundary restricts the interval of the
$k-$integration in Eq.\ \ref{sqw} to the vicinity of $k\!=\!\pm\!\pi$. This allows us to linearize the
bottom of the spectrum of the spin excitations writing
$\omega_s(k)\approx\pi J_{eff}(\pi-|k|)/2$ and to replace the
exponent $\alpha_k$ in Eq.\ \ref{cqw} by $\alpha\!=\!1/4$, its
value at the zone boundary. With this approximation the
fermionic correlator becomes :

\begin{eqnarray}
\label{alpha}
{\cal C}_k(q,\omega)\approx \theta(\omega) \frac{4 \pi}{k_F}
\frac{\exp\left(-\omega/E_F\right)}{\sqrt{2 q v_F}}
\frac{1}{\left(\omega^2-q^2 v_F^2\right)^{1/4}}\equiv {\cal
C}(q,\omega).
\end{eqnarray}

Carefully analyzing the contribution to the integral over $k$ in
Eq.\ \ref{sqw} of the different regions where the spectral
weights of $f$ and ${\cal C}$ are simultaneously large, and
using the fact that the structure factor varies smoothly
compared to the two other factors in the integral, we may write
the dynamic structure factor in the approximate form :

\begin{eqnarray}
\label{finalsqw}
{\cal S}(Q,\Omega)=|A_{+}|^2 \Phi(Q-\pi (1+c),\Omega)+|A_{-}|^2 \Phi(-Q+\pi (1-c),\Omega),
\end{eqnarray}

\noindent where we have defined

\begin{eqnarray}
\label{functionphi}
\Phi(\kappa,\Omega)=3a\int_0^{\Omega} \frac{d\omega\
e^{-\omega/E_F} }{v_s E_F (1+c)}\int_{\kappa}^{\infty} \frac{dq}{\pi
\sqrt{2|q|}}\left[\left(\frac{\Omega\!-\!\omega}{v_s}\right)^2
-|q\!-\!\kappa|^2\right]^{-1/2}
\left[\left(\frac{\omega}{v_F}\right)^2\!-\!q^ 2\right]^{-1/4}.
\end{eqnarray}

\noindent Here, $v_s=\pi J_{eff}/2(1+c)$ is the renormalized
spinwave velocity and the amplitudes $A_{\pm}$ are given by

\begin{equation}
\label{amplitudes}
A_{\pm}=1\pm\sin\frac{\pi c}{2}.
\end{equation}

Eq.\ \ref{finalsqw} contains the main result of this  work. The structure factor of the chains appears as the sum of two contributions. We will be seen below that $\Phi(\kappa,\Omega)$ peaks at $\kappa\sim 0$ and is exponentially small for $\kappa\ge0$. Therefore,  the first term is important for $Q\lesssim \pi(1+c)$ and the second one for $Q\gtrsim\pi(1-c)$. For vanishing hole concentration $A_{\pm}=1$, and the two $\Phi$ functions reduce to the $Q\le\pi$ and $Q\ge\pi$ parts of $f(Q,\Omega)$, respectively. For a finite hole concentration the corresponding peaks shift and broaden and the result is a distorted lineshape. 

In order to make further progress we must 
analize in detail the properties of $\Phi(\kappa,\Omega)$. We begin by examining the properties of the instantaneous
function obtained by integrating Eq.\ \ref{functionphi} over
frequency. The result can be expressed in terms of the
convolution of the two static functions defined in Eqs.\
\ref{fstatic} and \ref{cstatic},

\begin{eqnarray}
\label{phikappa}
\Phi(\kappa)=\frac{3a}{2\pi (1+c)} \int_{\kappa}^{\infty}
\frac{dq}{2\pi} {\cal C}(q) \log\left|\frac{4
(1+c)}{q-\kappa}\right|,
\end{eqnarray}

\noindent where ${\cal C}(q)\equiv {\cal C}_{\pi}(q)$ (cf. Eq.\
\ref{cstatic}) and we have used the asymptotic form of $f(k)$
given in Eq.\ \ref{fstatic}.

For $c\rightarrow 0$ and $c\rightarrow 1$ where $k_F$ vanishes,
${\cal C}(q)\rightarrow 2\pi  \delta(q)$ and $\Phi (\kappa)
\rightarrow 3a/(2\pi(1+c)) \theta (-\kappa)
\log\left|4(1+c)/\kappa\right|$. For intermediate values of the
hole concentration $\Phi (\kappa)$ may be obtained analytically
in two limits. At large $|\kappa/k_F|$ we find :

\begin{equation}
\label{large_kappa}
\Phi (\kappa)\sim \left\{
\begin{array}{ll}
\displaystyle
\frac{3a}{2\pi (1+c)} \log\left|\frac{4(1+c)}{\kappa}\right|
&, \kappa \ll -k_F\\
\displaystyle
\frac{3a k_F}{16\pi (1+c)^2}\ \frac{2^{1/4}}{\Gamma(1/4)}
\left(\frac{\kappa}{k_F}\right)^{-3/4} e^{-\kappa/k_F}\
\left(\gamma-1+\log\left[\frac{4(1+c)}{k_F}\right] \right)
&,  \kappa \gg k_F
\end{array}
\right.,
\end{equation}

\noindent where $\gamma$ is the Euler constant. We thus recover
the behavior of the pure system for large negative values of
$\kappa$.  The exponential tail that we find for large and
positive $\kappa$ results  from the smearing of the unperturbed
correlation function by the moving holes.

We can also evaluate analytically Eq.\ \ref{phikappa} for small
$\kappa/k_F$. The result is :

\begin{equation}
\label{small_kappa}
\Phi (\kappa)\sim \left\{
\begin{array}{ll}
\displaystyle
\Phi (0)-\frac{6a}{(2\pi^3)^{1/2}(1+c)}\ \sqrt{\kappa/k_F}\
\left(1+1/2\ \log\left|(1+c)/\kappa\right|\right)
&, \kappa>0\\
\displaystyle
\Phi (0)-\frac{6a}{(2\pi^3)^{1/2}(1+c)}\ \sqrt{|\kappa|/k_F}\
\left(\pi/2-1-1/2\ \log\left|(1+c)/\kappa\right|\right)\
&, \kappa<0
\end{array}
\right.,
\end{equation}

\noindent where $\Phi(0)$, the value at $\kappa=0$, is

\begin{eqnarray}
\label{zerok}
\Phi\left(0\right)=\frac{3a}{4\pi (1+c)}\ \left(\log\frac{4\
(1+c)}{k_F}\ +\ c_0\right),
\end{eqnarray}

\noindent and $c_0=\gamma+\pi/4+3/2 \log 2\approx\ 2.4$. For
a finite concentration of holes $\Phi (\kappa)$ has a cusp at
$\kappa\!=\!0$, a less singular behavior than at $c=0$ where
it diverges logarithmically at $\kappa=0$.

The two asymptotic forms, Eqs.\ \ref{large_kappa} and
\ref{small_kappa}, match smoothly at $\kappa \approx k_F$ as shown in Fig.  \ref{fig-static} where we plot the
$\kappa\!-$dependence of $\Phi (\kappa)$ for different hole
concentrations.

We now turn to a discussion of the frequency dependence of the
spectral function. The excitation spectrum of the pure model
goes down to zero-energy only at $\kappa\!=\!0$. The same is
true for a finite hole concentration. Putting $\kappa=0$ in Eq.\
\ref{functionphi} and performing the integral we find :

\begin{eqnarray}
\label{dynamicphi}
\Phi(0,\Omega)=\frac{1}{2 E_f \left(1+c\right)}
\int_{0}^{\left(1+v_s/v_F\right)^{-1}} \frac{dx}{1-x}\
\psi(x,\Omega)\  F\left(1/4,1/2,1,h^{-2}(x)\right)+
\nonumber
\\
+\frac{1}{2 k_F \sqrt{2\pi v_s v_F}}
\frac{\Gamma(1/4)}{\Gamma(3/4) (1+c)}
\int_{\left(1+v_s/v_F\right)^{-1}}^{1} \frac{dx}{\sqrt{x
(1-x)}}\ \psi(x,\Omega)\ F\left(1/4,1/4,3/4,h^{2}(x)\right),
\end{eqnarray}

\noindent where we have defined  the functions $\psi
(x,\Omega)\!=\!\theta
(x\!-\!1\!+\!\pi J_{eff}/\Omega)\ e^{-\Omega x/E_f}\!$,  and
$h(x)\!=\! (v_F/v_s) (1-x)/x$. Fig. \ref{fig-dynamic}
shows a plot of $\Phi(0,\Omega)$ as given by Eq.\
\ref{dynamicphi} for several hole concentrations and for
$\tau/J_{eff}=3$, a value consistent with our parametrization of
the Emery model.

In the pure Heisenberg model the spectral weight,
$\Phi(0,\Omega)\approx \Omega^{-1}$, diverges at zero frequency.
This divergence is cutoff for finite $c$ by the motion of the
holes and the intensity at $\Omega=0$ is now finite. The weight
missing at low energies is transferred to the high energy end
of the spectrum where an exponential tail,
$\Phi(0,\Omega)\!\sim\!\exp(-\Omega/E_F) (\pi J_{e
ff}/\Omega)^{-1/2}$, extends beyond the energy
$\Omega_{max}\!=\!\pi J_{eff}$ where the spectrum of the pure
Heisenberg model has a sharp cutoff. Analysis of Eq.\
\ref{functionphi} for finite $\kappa$ is more complicated.  In
this case the square roots present 
.in the integrand of  Eq.\
\ref{functionphi} impose the inequalities $v_F |q| \le \omega \le \Omega - v_s |q-\kappa|$ on the domain of integration. This condition can only be satisfied if $ \Omega \ge \kappa \min
(v_F, v_s )$, ${\it i.e.}$, there is a threshold in
$\Phi(\kappa,\Omega)$. Let us first consider the case $v_F>v_s$.
The bottom of the spinwave continuum starts at $\Omega=v_s
|\kappa|$. Near the onset, for $\Delta
\Omega\!\equiv\!\Omega-\Omega_{min} \ll \Omega$, only the
neighborhood of $q\!=\!0$ contributes to the integral in Eq.\
\ref{finalsqw}. Making the change of variables $q = (\omega/v_F)
\sin\theta$ the latter may be rewritten as :

\begin{eqnarray}
\Phi(Q,\Omega)&=&
\int_{0}^{\Omega}
\frac{d\omega \sqrt2 e^{-\omega/E_f}  }{ E_f(1+c)2\pi}
\int_{-\pi/2}^{\pi/2} \frac{d\theta}{\sqrt{|\tan\theta|}}
\left(\Delta\Omega + 2 \kappa v_s
-\omega(1+\frac{v_s}{v_F}\sin\theta) \right)^{-1/2}
\nonumber
     \\
     &&\times
     \left(\Delta\Omega
     -\omega(1-\frac{v_s}{v_F}\sin\theta) \right)^{-1/2}.
     \end{eqnarray}

\noindent  The first factor in the integral over $\theta$ is
dominated by $2\kappa v_s \simeq \Omega \gg \Delta\Omega$ which
allows us to approximate

     \begin{eqnarray}
      \nonumber
     \Phi(\kappa,\Omega)&\approx&
     \int_{0}^{\Omega}
     \frac{d\omega e^{-\omega/E_f}}{ E_f(1+c)2\pi
     \sqrt{\kappa v_s}}
     \int_{-\pi/2}^{\pi/2} \frac{d\theta}{\sqrt{|\tan\theta|}}
     \left(\Delta\Omega
     -\omega(1-\frac{v_s}{v_F}\sin\theta) \right)^{-1/2}
     \\&\simeq&
     \frac{\sqrt{\Delta\Omega}} { E_f(1+c)\pi
     \sqrt{\kappa v_s}} \int_{-\pi/2}^{\pi/2}
     \frac{d\theta}{\sqrt{|\tan\theta|}
     (1-\frac{v_s}{v_F}\sin\theta) }.
     \end{eqnarray}

\noindent The last integral can be done and we find the result

     \begin{equation}
     \label{largeVF}
     \Phi(\kappa,\Omega) \simeq
     2
     \left(\frac{\Delta\Omega}{2\kappa v_s} \right)^{1/2}
     \frac {\left(1-{v_s^2}/{v_F^2}\right)^{-1/4}}
     {E_f(1+c)}.
     \end{equation}

\noindent In the case $v_s>v_F$, the bottom of the excitation
spectrum is at $\Omega=v_F |\kappa|$, {\it i.e.}, below the
spinwave continuum. The integral over $q$ in  Eq.\
\ref{functionphi} is now dominated by $q\approx \kappa$. The
integration may be performed in a way similar to the previous
one and the result is  :

     \begin{equation}
     \label{smallVF}
     \Phi(\kappa,\Omega)\simeq
     \frac43
     \left(\frac{\Delta\Omega }{ 2\kappa v_F}\right)^{3/4}
     \frac{      \left( {v_s^2}/{v_F^2}-1\right)^{-1/2}      }
     { E_f(1+c)}         e^{-\kappa/k_f},
     \end{equation}

\noindent  The motion of the holes modifies substantially the
low energy part of the excitation spectrum of the chains. For
finite $c$, $\Phi(\kappa,\Omega)$ vanishes at threshold (cf.
Eqs.\ \ref{largeVF} and \ref{smallVF}) in contrast with the
situation at $c=0$ where the spectral weight has an inverse
 square root divergence at $\Omega=v_s |\kappa|$.  For
$c\!\approx\!0$ and $c\!\approx\!1$ where $v_F<v_s$, there are
hole-like excitations below the spinwave continuum. However,
their total weight is small because they are confined to the
region $|q-\pi(1+c)|\lesssim k_F$ by the exponential factor in
Eq.  \ \ref{smallVF}. Elsewhere in momentum space the spectrum
is essentially that of the Heisenberg antiferromagnetic chain.
At intermediate hole concentrations $v_F>v_s$ and the excitation
spectrum starts at the spinwave continuum but there exist
excitations with energies up to $E_F$ that drain spectral weight
from the region $\omega\lesssim J_{eff}$. The crossover between
these two regimes takes place when $v_s=v_F$, which for our
parametrization of the model yields a critical concentration
$c_{crit} \approx \frac1\pi \sin^{-1} (\pi J'/4\tau) \simeq 0.06$.

\section{discussion}
\label{sec:discussion}

The natural tool for experimental investigation of the magnetic
fluctuations described above is inelastic scattering of spin
polarized neutrons.  In this Section we will use our results
to compute the differential cross-section for
magnetic neutron scattering. The latter is related to the dynamic
structure factor of Eq.\ \ref{sqw} by the expression
\cite{lovesey}

\begin{eqnarray}
     \frac{d\sigma(q,\omega)}{d\Omega_s d\omega} &=&
     \frac{2}{3} (r_0\gamma)^2
     \frac{k_f} { k_i}
     {\cal S}(q,\omega),
     \label{crosssection}
\end{eqnarray}

\noindent where $r_0=e^2/mc^2$, and $\gamma= 1.57$ is the
neutron's gyromagnetic ratio. The neutron's initial and final
momenta are $k_i$ and $k_f$, respectively, $q$ is the momentum
transfer in the direction parallel to the chain, $\omega$ is the
neutron 's energy loss and $d\Omega_s$ is the element of solid
angle in the direction of the outgoing neutron.

The intensity scattered by a quasi one-dimensional object is uniformly distributed in a plane perpendicular to it. Therefore, the intensity at any particular point in $q-$space is very small and it is generally impossible to resolve the structure factor both in $q$ and $\omega$. Under typical experimental conditions what is measured is the cross-section integrated in energy from $0$ up to the incident neutron energy, $\omega_N$. If the experiment is performed under quasielastic conditions ({\it i.e.}, if most of the spectral weight is at energies smaller than $\omega_N$) we have

\begin{equation}
\label{quasielastic}
\frac{d\sigma(q)}{d\Omega_{\bf q}}=\frac{2}{3} (r_0\gamma)^2
{\cal S}(q)= 194\ {\text {mbarn}}\ {\cal S}(q),
\end{equation}

\noindent where is has been assumed that $\int_{0}^{\omega_N}
d\omega\ {\cal S}(q,\omega)\approx \int_{0}^{\infty} d\omega\
{\cal S}(q,\omega)={\cal S}(q)$.

Using the results of the preceding Section we may write

\begin{eqnarray}
\label{structure}
{\cal S}(q)=W_{+} \Phi(q-\pi (1+c))+W_{-} \Phi(-q+\pi (1-c)).
\end{eqnarray}

\noindent where $W_{\pm}=A_{\pm}^2$, and $\Phi(\kappa)$ was
defined in Eq.\ \ref{phikappa}. Figure \ref{fig-inten} shows the
evolution of ${\cal S}(q)$ with doping. In the limiting cases
$c\!=\!0$ and $c\!=\!1$ the system is equivalent to a
Heisenberg antiferromagnet. In the undoped case,
$W_{+}\!=\!W_{-}\!=\!1$ and the two terms in Eq. \ref{structure}
contribute to the peak at $q\!=\!\pi$, one for $q\!\leq\!\pi$,
the other for $q\!\geq\!\pi$ (cf. discussion below  Eq.\
\ref{amplitudes}). In the fully doped case $W_{+}\!=\!4$,
$W_{-}\!=\!0$ and only the first term remains giving a peak at
$q\!=\!2\pi$. The fourfold increase in the amplitude $W_{+}$
that occurs when going from $c\!=\!0$ to $c\!=\!1$ is halved by
the factor $(1+c)^{-1}$ that appears as a prefactor in the
expression for $\Phi(\kappa)$.  Therefore, the area
$\int_{0}^{2\pi} dq/(2\pi)\ S(q)$, proportional to the number of spins
per unit cell, just doubles as it must. As the hole
concentration increases from $c\!=\!0$ to $c\!=\!0.5$ the peak
shifts and the lineshape becomes increasingly different from to
that at of the Heisenberg chain because the interval of momenta
in which the presence of holes is relevant, $\Delta q\approx
k_F$, gets bigger. Upon further increase of $c$ the process
reverts because $k_F $ decreases back to zero (cf. the paragraph
above Eq. \ref{defrho}).

The hypothesis of quasielastic scattering implicit in the precceding discussion is unlikely to be fulfilled under realistic experimental conditions. For $c\sim 0.3$, the excitation spectrum extends up to energies of the order of $E_F\sim 1\,$eV,  whereas $\omega_N\lesssim 50\,$ meV in a typical spin-polarized neutron scattering experiment. An important reduction of intensity is thus expected to result from the fact that the excitation spectrum is only partially integrated. The peak intensity can be estimated by integrating Eq.\ \ref{crosssection} at $q\!=\!\pi(1+c)$ up to $\omega_N$, using Eq.\ \ref{dynamicphi}. We find $d\sigma/d\Omega_s|_{max}\!\approx 60\,$mbarn for $\omega_N\!\approx\!50\,$mev whereas, under the same conditions,  Eq.\ \ref{quasielastic} would predict $d\sigma/d\Omega_s|_{max}\!\approx 800\,$mbarn. However, despite the large decrease of intensity, the predicted signal might still be observable in a high flux reactor.

We conclude by examining the expected finite-temperature corrections to the results obtained in this paper.
The effects of temperature are different in the fermionic and spin
subsystems.  We show in Appendix \ref{sec:app:temp} that the
power-law behavior of the correlation function ${\cal
C}_k(l;t=0)$ that we found in Eq.\ \ref{defc}, gives way to 
exponential decay at finite $T$ :

\begin{eqnarray}
{\cal C}_k(l;t=0)= \exp\left(-\frac{k^2 l}{\pi\xi}\right),
\end{eqnarray}

\noindent where the typical scale of the correlations is
$\xi\!=\!v_F/T$.  The appearance of a finite decay length
rounds-off the singularities of $C(q,\omega)$ at a scale $q\sim
\xi^{-1}$, $\omega\!\sim\!T$. At room temperature $T\simeq
300\,$K,  and for  $v_F\sim 1\,$eV this should lead to a
smearing of the peaks in Fig.\ \ref{fig-inten} on a scale $\sim
0.03\,$r.l.u., which is of the order of magnitude of the
experimental resolution.

The changes in the spectral properties of the Heisenberg
chain at finite $T$ have been discussed by M\"uller et al.
\cite{MTBB} who have shown that spectral weight
appears below the bottom of the spin-wave continuum with
increasing the temperature. At $k_B T/J = 0.5$, the total
intensity in the region $\omega < \omega_s(k)$ , is comparable
to that at $\omega > \omega_s(k)$ but for $k_B T/J\simeq 0.1$
these effects are negligible and the correlation function is
essentially identical to that at $T=0$. For CuO chains where
$J_{eff}\simeq 0.15\,$eV and if the experiment is performed just
 above the superconducting transition temperature at $T\simeq
100\,$K, the ratio is $T/J\simeq 0.1$. Under these conditions the predictions of our $T=0$ calculations should remain valid.

\acknowledgements
We thank V.P. Plakhty for useful discussions on  the
experimental aspects of this work.  One of us (D.N.A.)
acknowledges financial support from
the Russian Foundation for Basic Research ( Grant No.
96-02-18037-a ) and a PECO fellowship from the French Minist\`{e}re de la Recherche.


\appendix

\section{the coherence factor}
\label{sec:app:coh}

In this Appendix we derive the form of the coherence factor
$A(Q,k)$ defined in Eq.\ \ref{coherence}. The Fourier components
of the spin density are given by \cite{foot2}

\begin{equation}
\label{fourier}
{\bf S}_Q=\sum_l e^{-iQl} \left({\bf S}_l+e^{-iQ/2} {\bf
S}_{l+1/2}\right)\equiv \left.{\bf S}_Q\right|_{Cu}+\left.{\bf
S}_Q\right|_{O},
\end{equation}

\noindent where the right hand side of the equation defines the
Cu and O contributions to ${\bf S}_Q$.  Using the expansions
\ref{spinop1} and \ref{spinop2} above and the identity
$\hat{n}=\left(\exp ik\hat{n}-1\right)/\left(\exp ik-1\right)$
valid for an y number operator $\hat{n}$ and any value of $k$,
the oxygen contribution may be cast in the form :

\begin{equation}
\label{o_fourier}
\left.{\bf S}_Q\right|_{O}=(N+N_h)^{-1/2} \sum_k {\bbox
\sigma}_k \sum_l \exp i\left[(k-Q)(l+1/2)+k \varphi
(l)\right]\frac{\exp ik\hat{n}_{l+1/2}-1}{2i\sin(k/2)},
\end{equation}

\noindent where $\varphi (l)=\sum_{j=0}^{l-1} \hat{n}_{j+1/2}$.
Noticing that $\varphi (l)+\hat{n}_{l+1/2}=\varphi (l+1)$ this
is equivalent to

\begin{equation}
\label{o_fourier1}
\left.{\bf S}_Q\right|_{O}=(N+N_h)^{-1/2} \sum_k {\bbox
\sigma}_k \sum_l \exp i(k-Q)(l+1/2) \frac{\exp
ik\varphi(l+1)-\exp i k\varphi(l)}{2i\sin(k/2)}.
\end{equation}

Changing indices of summation from $l$ to $l'=l+1$ in the first
summation in the equation above and recalling that periodic
boundary conditions impose that $\varphi(N)-\varphi(0)=N_h$ and
that $kN(1+c)=2\pi m_1$ and $QN=2\pi m_2$ with $m_1$ and $m_2$
integers, we can rewrite Eq.\ \ref{o_fourier1} in the form

\begin{equation}
\label{o_fourier2}
\left.{\bf S}_Q\right|_{O}=(N+N_h)^{-1/2} \sum_k {\bbox
\sigma}_k \frac{\sin((Q-k)/2)}{\sin(k/2)} \sum_l \exp i[(k-Q)
l+k \varphi(l)].
\end{equation}

Adding to this expression the trivial Cu contribution we get :

\begin{equation}
\label{fourier2}
{\bf S}_Q=(N+N_h)^{-1/2} \sum_k {\bbox \sigma}_k A(Q,k) \sum_l
\exp i((k-Q) l+k \varphi(l)),
\end{equation}

\noindent where we defined the coherence factor

\begin{equation}
\label{result}
A(Q,k)=1+\frac{\sin((Q-k)/2))}{\sin(k/2)}.
\end{equation}

\section{evaluation of the fermionic correlator}
\label{sec:app:fermionic}

The fermionic correlator is given by the expression

\begin{eqnarray}
\label{app1}
\left<\exp ik{\cal A}(l,t) \ \exp -ik
{\cal A}(l',t')\right>_{H_1},
\end{eqnarray}

\noindent where the bosonic representation of ${\cal A}(l,t)$
has been given in Eq.\ \ref{defa} above. Eq.\ \ref{app1} may be
easily evaluated by repeated application of the well-known
formula

\begin{eqnarray}
\label{baker}
\exp(A+B)=\exp A \exp B \exp(-1/2[A,B]),
\end{eqnarray}

\noindent valid for any pair of operators $A$ and $B$ that
satisfy $[A,[A,B]]=[B,[A,B]]=0$.

We first use Eq.\ \ref{baker} in Eq.\ \ref{app1} in order to
combine the two exponentials into one,

\begin{eqnarray}
\label{app2}
\begin{array}{ccc}
\left<\exp ik{\cal A}(l,t) \ \exp -ik
{\cal A}(l',t')\right>_{H_1}=\\
=\left<\exp ik \left\{{\cal A}(l,t)-{\cal A}(l',t')\right\}\right>_{H_1}  \exp \left\{\frac{ k^2}2
\left[{\cal A}(l,t),{\cal A}(l',t')\right]\right\},
\end{array}
\end{eqnarray}

\noindent where the commutator on the right hand side of Eq.\
\ref{app2} is a $c-$number given by (cf. Eq.\ \ref{defa}) :

\begin{eqnarray}
\label{commutator1}
\left[A(l,t),A(l',t')\right]=- 2 i \sum_{q} v^{2}(q) \sin
\left(v_F |q| (t-t')\right) Re\left[\phi(q,l)
\phi(-q,l')\right].
\end{eqnarray}

\noindent Next, we write
${\cal A}(l,t)-{\cal A}(l',t')$ as the difference of two bosonic operators,

\begin{eqnarray}
\label{difference}
\begin{array}{ccc}
{\cal A}(l,t)-{\cal A}(l',t')={\cal B}(l,l';t,t')+{\cal
B}^{\dagger}(l,l';t,t'),\\
{\cal B}(l,l';t,t')=\sum_q v(q) \left(\phi(q,l) e^{-i \omega_q
t}-\phi(q,l') e^{-i \omega_q t'}\right) b_{-q}^{ }.
\end{array}
\end{eqnarray}

\noindent Using once more Eq.\ \ref{baker} we obtain :

\begin{eqnarray}
\label{bosongs}
\langle\exp ik({\cal B}+{\cal B}^{\dagger})\rangle
=\exp-\frac{k^2}2 [{\cal B},{\cal B}^{\dagger}],
\end{eqnarray}

\noindent where

\begin{eqnarray}
\label{commutator2}
[{\cal B},{\cal B}^{\dagger}]=2 \sum_{q} v^{2}(q) \left\{\
\frac{{\left|\phi(q,l)\right|}^2+{\left|\phi(q,m)
\right|}^2}{2}-\cos \left(v_F |q| t\right)
Re\left[\phi(q,l)\phi(-q,m)\right]\right\},
\end{eqnarray}

\noindent and we have used the fact that no bosons are excited
in the ground-state. Combination of these results leads to the
expression quoted in Eq. \ref{corrfermions} of the main text.

\section{temperature dependence of the fermionic correlator}
\label{sec:app:temp}

We show below that the correlation length of the fermion
subsystem at finite temperature is $\xi=v_F/T$.  For this we
use a modification of an approach originally due to Lieb,
Schultz and Mattis \cite{LSM}. It is based on a representation
of the instantaneous correlation function ${\cal C}_Q(l,0;t=t')$
of Eq.\ \ref{fermionic} as the Toeplitz determinant

     \begin{eqnarray}
     \label{toeplitz}
     {\cal C}_Q(l)& =& \langle
     \exp{iQ \sum_q\Phi(q,l)\rho_q} \rangle
     = det(M)                          \\
     M_{kl}& = &\delta_{kl} - (1-e^{iQ}) g_{kl}.
     \nonumber
     \end{eqnarray}

The $l\!\times\!l$ matrix $M$ is constructed out the fermionic
correlation functions $g_{km} = \sin(\pi c(k-m) ) /\pi(k-m) =
(1/2\pi) \int_{-\pi}^\pi dq\, n_q e^{iq(k-m)}$ where the Fermi
function $n_q =1/ (1+e^{\varepsilon_q/T})$.  Notice that here we
do not subtract the $q=0$ component of the density in Eq.\
\ref{toeplitz}. Next, we use the asymptotic behavior of the
above determinant at large $l$,

\begin{equation}
\label{determinant}
     det(M) \sim \exp[l D - K_2],
\end{equation}

\noindent where \cite{Luttinger}

     \begin{eqnarray}
     D = iQ \int_{-\pi}^\pi \frac{dq}{2\pi} \tilde n_q,
     \end{eqnarray}

     \begin{eqnarray}
     K_2 =  Q^2 \int_{-\pi}^\pi \frac{dq\,dk}{2(2\pi)^2}
     {\tilde n}_q ( 1- {\tilde n}_{q+k})
     \frac{1-\cos kl}{1-\cos k},
     \label{app:k2}
     \end{eqnarray}

\noindent and $iQ {\tilde n}_q = \ln[1 - (1-e^{iQ}) n_q ]$.

At $T=0$ ${\tilde n}_q = n_q $ and we recover our results in the
main text. For finite temperature, ${\tilde n}_q $ is not so
simple. For our purposes it is sufficient to expand ${\tilde
n}_q$ up to terms of order $Q^2$ :

\begin{equation}
\label{expansion}
     iQ {\tilde n}_q = iQ n_q - \frac{Q^2}2 n_q(1-n_q),
\end{equation}

\noindent from which the corrections to the two terms that
appear in the exponent in Eq.\ \ref{determinant} follow :

     \begin{equation}
     lD = iQ cl - \frac{Q^2l}{2\pi\xi},
     \label{app:DT}
     \end{equation}

     \begin{eqnarray}
     K_2(T)-K_2(0) &=&
     \frac {Q^2}{2\pi^2} \int_0^\pi
     \frac{dk} k \,
     \frac{1-\cos kl}{\exp(kv_F/T)-1}
     \nonumber\\
     & = &
     \frac {Q^2}{2\pi^2} \ln\frac{\sinh\pi l/\xi}
     {\pi l/\xi}.
     \label{app:k2T}
     \end{eqnarray}

Combining  (\ref{app:DT}) and  (\ref{app:k2T}), one finds that
at $Q\sim\pi$ and large $l$

\begin{equation}
\label{result1}
     |{\cal C}(l)| \sim
     \exp\left[ -\pi l\xi \right].
\end{equation}

\begin{figure}
\caption{The function $\Phi(\kappa)$
for the CuO chains for several values of the hole concentration $c$. }
\label{fig-static}
\end{figure}

\begin{figure}
\caption{Frequency dependence of $\Phi(0,\Omega)$ for three values of the hole
concentration $c$. The ratio of the effective exchange constant to the hole hopping integral is $J_{eff}/\tau\!=\!1/3.
$}
\label{fig-dynamic}
\end{figure}

\begin{figure}
\caption{
Equal-time spin-spin correlation of the CuO chains as function of
q and of the hole concentration $c$.
}
\label{fig-inten}
\end{figure}

%
%
%


\begin{references}
\bibitem{foot1} For finite $U_d$ there are hopping processes in which the spin sequence changes but their amplitude, proportional to $t^2/(U_d-\epsilon_p)$, vanishes in the limit considered here.
\bibitem{foot2} We take the form factors of the Cu$^{+2}$ an
O$^{-1}$ ions equal to unity, an excellent approximation in the
domain of wavelength typical for magnetic neutron scattering.
\end{references}
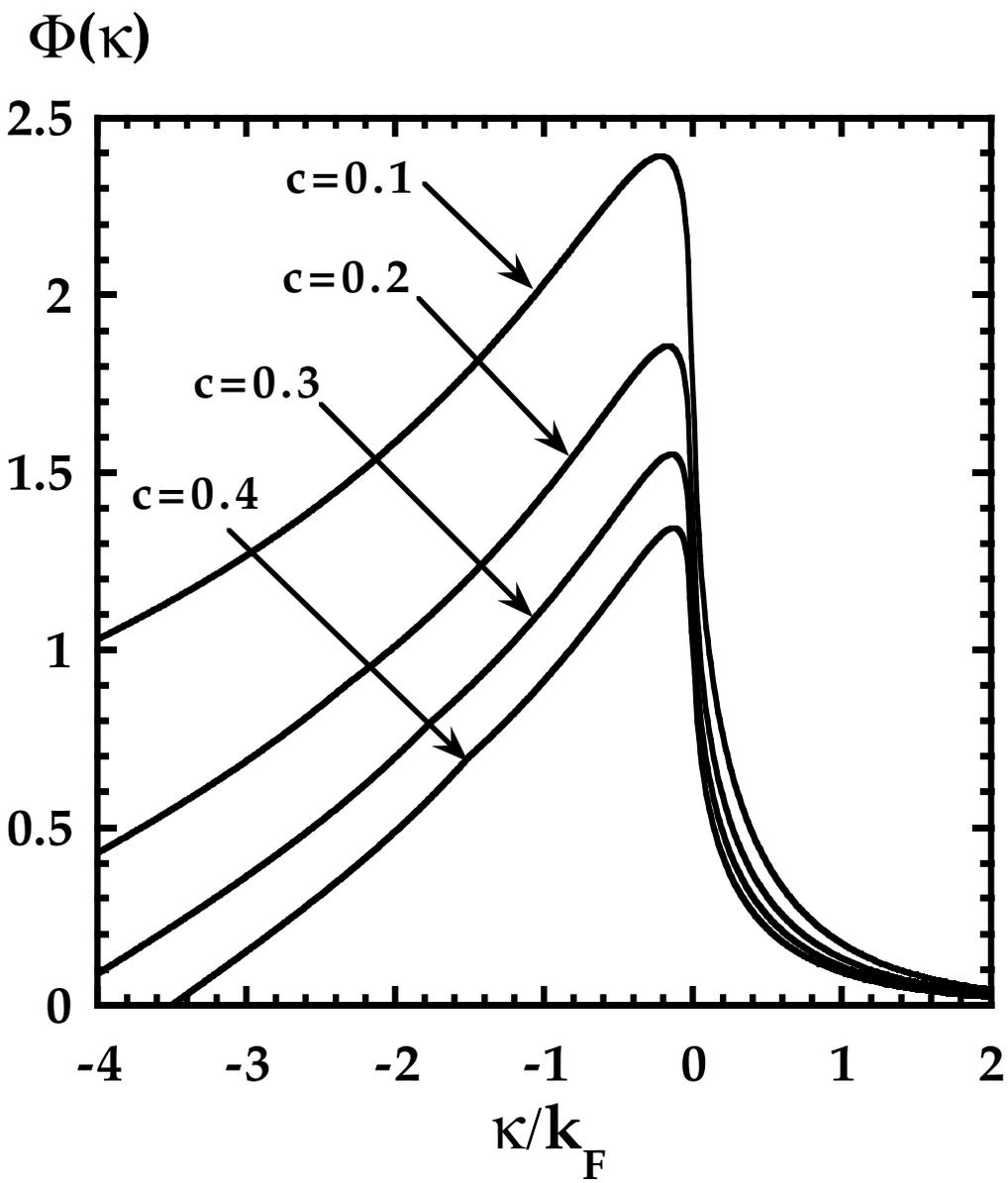
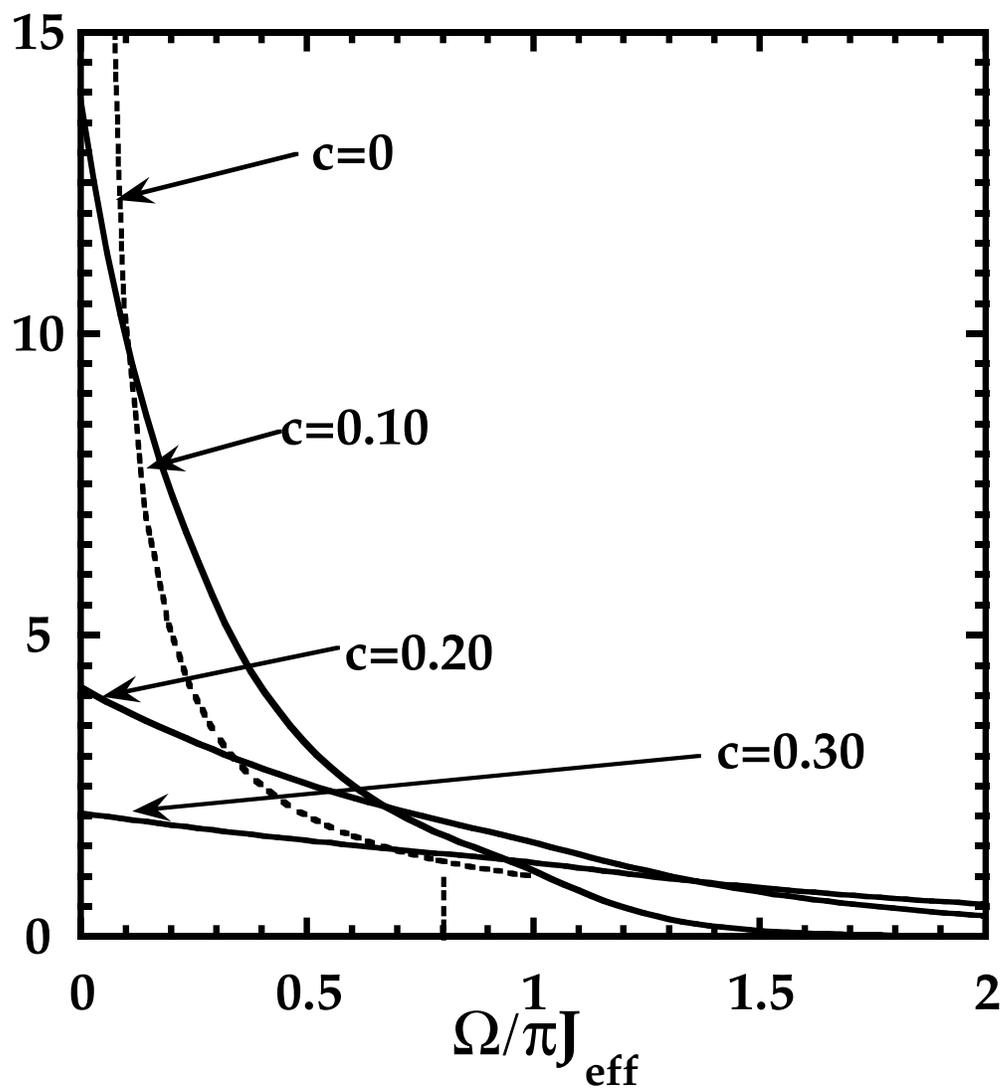
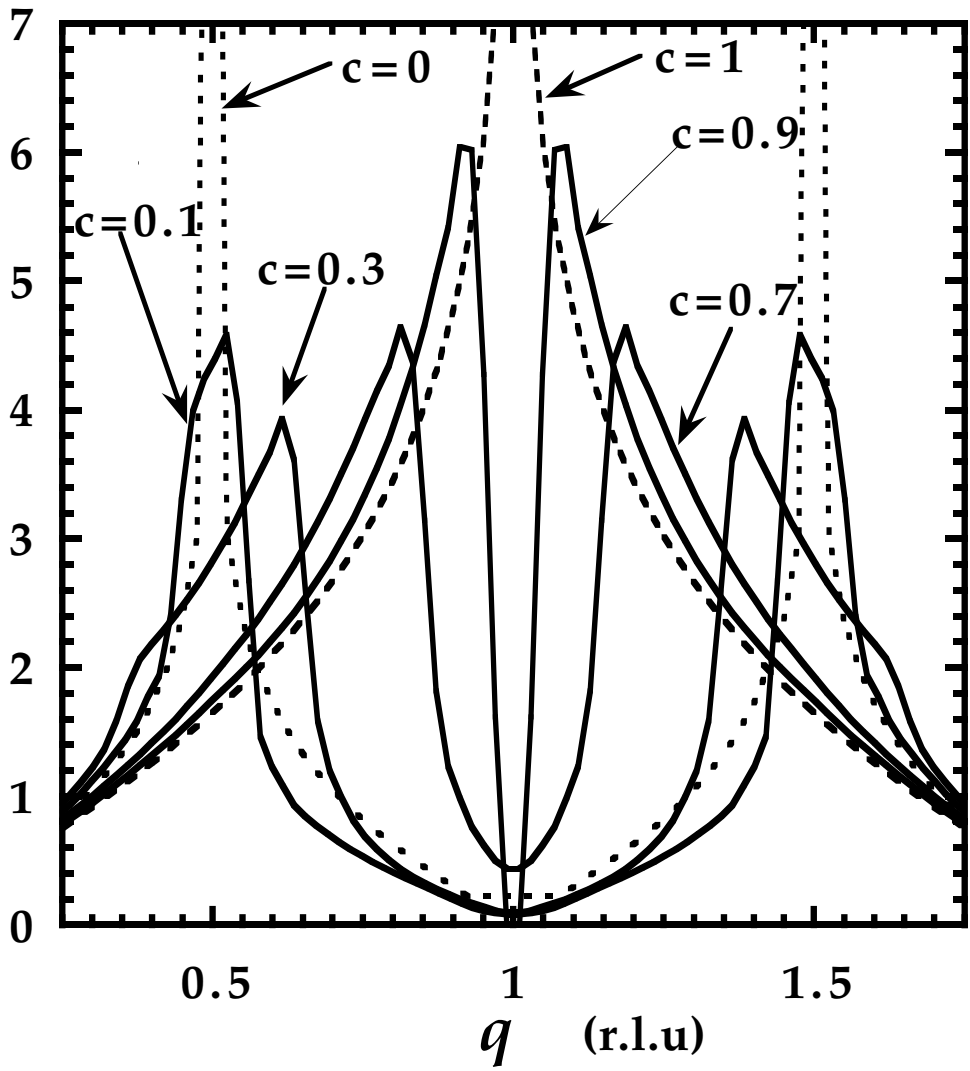
\end{document}